\documentclass[runningheads]{llncs}
\usepackage[T1]{fontenc}
\usepackage{graphicx}
\usepackage{hyperref}
\begin{document}
\title{Sensitivity-Aware Retrieval-Augmented Intent Clarification}
%
%
\author{Maik Larooij\inst{1}\orcidID{0000-0001-9616-1666}}
\authorrunning{M. Larooij}
%
\institute{ICAI OpenGov Lab x IRLab, University of Amsterdam, The Netherlands\\
\email{m.k.larooij@uva.nl}}
\maketitle              
\begin{abstract}
In conversational search systems, a key component is to determine and clarify the intent behind complex queries. We view intent clarification in light of the \textit{exploratory search} paradigm, where users, through an iterative, evolving process of selection, exploration and retrieval, transform a visceral or conscious need into a formalized one. Augmenting the clarification component with a retrieval step (\textit{retrieval-augmented intent clarification}) can seriously enhance clarification performance, especially in domains where Large Language Models (LLMs) lack parametric knowledge. However, in more sensitive domains, such as healthcare, government (e.g. FOIA search) or legal contexts, the retrieval database may contain sensitive information that needs protection. In this paper, we explore the research challenge of developing a retrieval-augmented conversational agent that can act as a mediator and gatekeeper for the sensitive collection. To do that, we also need to know what we are protecting and against what. We propose to tackle this research challenge in three steps: 1) define an attack model, 2) design sensitivity-aware defenses on the retrieval level and 3) develop evaluation methods to measure the trade-off between the level of protection and the system's utility.

\end{abstract}
\section{Introduction}
Recently, the field of information retrieval has experienced a paradigm shift from traditional search engines to conversational search interfaces \cite{mo2025survey,radlinski2017theoretical}. Instead of filtering and evaluating a ranked list of search results, these interfaces, nowadays powered by Large Language Models (LLMs), provide users with direct, coherent and context-aware answers in natural language, mimicking human dialogue. A crucial step in conversational search is understanding and clarifying the user's intent, for example by asking clarifying questions \cite{aliannejadi2020convai3,zamani2020generating,braslavski2017you}. As it may be hard for users to formulate their complex information need into a single query, the system can decide to ask users questions to clarify their intent when the question is too ambiguous, faceted or broad \cite{aliannejadi2019asking}. In recent work, instead of choosing from a pool of candidate questions, LLMs enable these systems to directly generate the next clarification step \cite{sekulic2021towards}.

We view intent clarification as a critical component of the exploratory search paradigm. Compared to `lookup' search (question answering, fact retrieval, known item search), users engage in exploratory search to learn or investigate \cite{marchionini2006exploratory}. At the lowest level, exploratory search may start with a visceral, inexpressible need, a vague sort of dissatisfaction, also coined an `anomalous state of knowledge' \cite{belkin1980anomalous}. Through an iterative process of selection, exploration and retrieval, users form a focus based on the information encountered and are then able to formally express their information need \cite{kuhlthau1991inside,taylor1967question}. Bates \cite{bates1989design} described this as `evolving search', in which the query itself is continuously changing, by collecting bits of information along the way, analogous to `berrypicking'.  

Previous studies have shown that retrieving relevant context, such as early result sets, can be beneficial for intent detection and clarification, for example in query expansion and pseudo-relevance feedback \cite{buckley1995automatic,cao2008selecting}, intent classification \cite{srinivasan2022quill,zhang2020query} and clarifying questions \cite{mass2022conversational,krasakis2024corpus}. This \textit{retrieval-augmented intent clarification} is especially important in situations with domain-specific information for which LLMs lack sufficient parametric knowledge. These domains, however, are sometimes sensitive or regulated, such as government, healthcare or legal contexts. Context information may live in private databases, not yet processed or publicly available. Systems must be aware of the potential leakage of sensitive information \cite{olteanu2021facts}. This problem has previously led to research on privilege review in e-discovery \cite{vinjumur2015finding,oard2018jointly}, technology assisted sensitivity review \cite{mcdonald2018active,mcdonald2021framework} and sensitivity-aware search (SAS) \cite{mckechnie2024bi,mckechnie2024cascading}. The sensitivity-aware use of information for LLM-based, retrieval-augmented intent clarification poses a challenge, as LLMs have been known to be able to leak various kinds of information and can be tricked into ignoring system instructions, an attack known as  `jailbreaking' \cite{zhou2025don,wei2023jailbroken,russinovich2025great}. 

In this paper, we set out to define the research challenge described above and propose a vision for sensitivity-aware, retrieval-augmented intent clarification agents that protect sensitive information in documents that should not be exposed in an exploratory (rather than lookup) clarification scenario. We show that the conversational agent must act as a mediator and gatekeeper between the user and sensitive document collection. 

\section{Research Challenge}
Imagine a librarian with extensive knowledge of the library's book collection. A visitor comes in, wanting to discover new books on history. The librarian, helpful as they are, knows that there are thousands of history books and asks: "What time period of history are you interested in?" "Before Christ," the visitor answers. The librarian, narrowing down the potential books in their mind, knows that there are many popular books on Greeks or Romans and proceeds: "Are you looking for books on Greeks or Romans?" The visitor hesitates and then says: "Now that you say this, I remember a Babylonian king I wanted to know more about". The librarian now knows enough and shows the visitor to the section on Babylonia. This example resembles the process of exploratory search. The librarian and visitor `negotiated' an unclear need into a formalized one, with the librarian acting as a \textit{mediator} between the visitor and the book collection.

Let's now consider another scenario. Several countries have legislation granting citizens the right to file information requests to the government, such as the Freedom of Information Act (FOIA) in the United States and the Wet open overheid (Open Government Act) in the Netherlands. Government officials, having knowledge of the available document collection, often proactively reach out to requesters to clarify or narrow down the scope of the request \cite{DOJ2025ChiefFOIA}, similarly acting as a mediator between the requester and the document collection. 

The key difference between these two scenarios is that, unlike in a library, governmental documents may contain sensitive information, prohibiting (partial) disclosure. The mediator in these situations is therefore also a \textit{sensitivity-aware gatekeeper} who decides which information may be shared with the requester.

Suppose we want to automate this exploratory clarification process and build a generative conversational agent that can act as both a mediator and a sensitivity-aware gatekeeper to jointly clarify the search intent between a user and a sensitive set of documents. However, unlike a domain expert, a generic LLM may not have a good and reliable grasp of what information must be protected. Also, LLMs do not excel at keeping secrets. Like other machine learning models, they have been found to be prone to Membership Inference Attacks (MIA) targeted at leaking training data \cite{duan2024membership,feng2024exposing,wu2025membership,kandpal2023user,galli2024noisy}, and they can be tricked into ignoring system instructions and guardrails (jailbreaking) \cite{zhou2025don,wei2023jailbroken,russinovich2025great}. 

There have also been MIAs targeting Retrieval-Augmented Generation (RAG), aimed at inferring whether a piece of text was retrieved from a private database and used for answer generation \cite{naseh2025riddle,li2025generating,anderson2025my,liu2025mask}. While these are closely related, as retrieval-augmented intent clarification can be viewed as an implementation of RAG, we argue that this exploratory process substantially differs from the generic Q\&A (lookup) RAG pipeline and raises some interesting new challenges. Membership inference attacks on RAG systems detect leakage through retrieved content and generated answers, for example by explicitly asking about membership \cite{anderson2025my}, letting the system fill in masked words \cite{liu2025mask} or carefully crafting very specific questions that can only be answered if the target document is present \cite{naseh2025riddle}. Compared to these `direct' attacks, an attack on the conversational intent clarification system cannot operate by directly asking or completing sentences. Instead, it must use indirect signals based on what the system asks rather than what it answers.



\section{Towards Sensitivity-Aware Intent Clarification}
To tackle the research challenge defined above, we need to work through three steps, described below.

\textbf{An Attack Model.}
First, a clear definition of the attack model must be created. This includes the attacker's goal, knowledge and capabilities, as well as the full setup of the intent clarification system. This also requires a very careful description of \textit{what} exactly is considered sensitive information. Sensitivity can exist at different levels, such as individual passages of text, whole documents, or entire collections. A comprehensive study should define the attack and the granularity of sensitive information, and measure the effectiveness of the attack.

\textbf{Retrieval-Based Sensitivity-Aware Defenses.}
Given a clear attack scenario, we can consider sensitivity-aware defenses. Existing tested or proposed defenses against MIA on RAG rely mostly on anomaly detection and guardrails in system prompts \cite{li2025generating,anderson2025my}. We argue that relying on the LLM itself for defense is not future-proof and results in a cat-and-mouse game with the attacker. Instead, we propose to study new defenses on the \textbf{retrieval level}. 

Previously, two paradigms, `protect-then-search' and `search-then-protect' have been proposed. The former means preprocessing the information into sensitivity-aware formats before searching, such as with technology-assisted sensitivity review \cite{mcdonald2021framework}, privacy-preserving text sanitization \cite{anandan2012t,sanchez2016c} or automatic FOIA redaction \cite{baron2023using}. We propose another interesting direction, inspired by the notion of \textit{k-anonymity} \cite{sweeney2002k}. This means creating abstractions of the documents - for example into topics, sentences or labels - to ensure that each document is indistinguishable from at least $k$ other documents. The `search-then-protect' approach relies on making the collection entirely accessible, hiding sensitive information when the system comes across it. An example of this is sensitivity-aware search \cite{mckechnie2024cascading}. We also propose a new approach here, inspired by the notion of \textit{differential privacy} \cite{dwork2006differential}: adding noise to the retrieval results to add uncertainty about the membership of documents in the collection. We argue that the added noise may be acceptable in situations that do not rely on directly outputting factual information, such as clarifying questions.

\textbf{Evaluation of Sensitivity-Aware Interventions.}
To evaluate the interventions, we need to define new evaluation methods that measure the trade-off between the level of protection and the utility of the system. For protection, we can use the success rate of attacks and rely on the privacy guarantees of the interventions and measure the utility at multiple privacy budgets. For utility, we propose to focus on the effect of the intent clarification process on a downstream task, such as relevant document retrieval. Two potential datasets with annotations on sensitivity and relevance are Avocado \cite{sayed2020test} and SARA \cite{mckechnie2024sara}.

\section{Conclusion}
This paper has set out a research challenge to develop sensitivity-aware interventions on top of a generative conversational agent, functioning as a mediator and gatekeeper between the user and a private document collection, to benefit from retrieval-augmented exploratory intent clarification while aiming to protect information in documents that should not be exposed. 

We propose to tackle this research challenge in three steps: 1) provide a clear definition of the attack model, 2) design sensitivity-aware interventions on the retrieval level and 3) develop evaluation methods to measure the trade-off between the level of protection and the utility of the system.
%
%
%
\bibliographystyle{splncs04}
\bibliography{refs}

@inproceedings{radlinski2017theoretical,
  title={A theoretical framework for conversational search},
  author={Radlinski, Filip and Craswell, Nick},
  booktitle={Proc CHIIR'17},
  pages={117--126},
  year={2017}
}

@article{mo2025survey,
  title={A survey of conversational search},
  author={Mo, Fengran and Mao, Kelong and Zhao, Ziliang and Qian, Hongjin and Chen, Haonan and Cheng, Yiruo and Li, Xiaoxi and Zhu, Yutao and Dou, Zhicheng and Nie, Jian-Yun},
  journal={ACM Transactions on Information Systems},
  volume={43},
  number={6},
  pages={1--50},
  year={2025},
  publisher={ACM New York, NY}
}

@inproceedings{zamani2020generating,
  title={Generating clarifying questions for information retrieval},
  author={Zamani, Hamed and Dumais, Susan and Craswell, Nick and Bennett, Paul and Lueck, Gord},
  booktitle={Proc WebConf'20},
  pages={418--428},
  year={2020}
}

@inproceedings{aliannejadi2019asking,
  title={Asking clarifying questions in open-domain information-seeking conversations},
  author={Aliannejadi, Mohammad and Zamani, Hamed and Crestani, Fabio and Croft, W Bruce},
  booktitle={Proc SIGIR'19},
  pages={475--484},
  year={2019}
}

@inproceedings{braslavski2017you,
  title={What do you mean exactly? Analyzing clarification questions in CQA},
  author={Braslavski, Pavel and Savenkov, Denis and Agichtein, Eugene and Dubatovka, Alina},
  booktitle={Proc. CHIIR'17},
  pages={345--348},
  year={2017}
}

@article{aliannejadi2020convai3,
  title={ConvAI3: Generating clarifying questions for open-domain dialogue systems (ClariQ)},
  author={Aliannejadi, Mohammad and Kiseleva, Julia and Chuklin, Aleksandr and Dalton, Jeff and Burtsev, Mikhail},
  journal={arXiv preprint arXiv:2009.11352},
  year={2020}
}

@report{DOJ2025ChiefFOIA,
  author       = {{United States Department of Justice}},
  title        = {Chief FOIA Officer Report},
  institution  = {U.S. Department of Justice},
  year         = {2025},
  type         = {Government Report},
  url          = {https://www.justice.gov/oip/united-states-department-justice-2025-chief-foia-officer-report},
  note         = {Accessed January 2026}
}

@inproceedings{kandpal2023user,
  title={User inference attacks on LLMs},
  author={Kandpal, Nikhil and Pillutla, Krishna and Oprea, Alina and Kairouz, Peter and Choquette-Choo, Christopher and Xu, Zheng},
  booktitle={Socially responsible language modelling research},
  year={2023}
}

@inproceedings{feng2024exposing,
  title={Exposing Privacy Gaps: Membership Inference Attack on Preference Data for LLM Alignment},
  author={Feng, Qizhang and Kasa, Siva Rajesh and KASA, SANTHOSH KUMAR and Yun, Hyokun and Teo, Choon Hui and Bodapati, Sravan Babu},
  booktitle={International Conference on Artificial Intelligence and Statistics},
  pages={5221--5229},
  year={2025},
  organization={PMLR}
}

@article{duan2024membership,
  title={Do membership inference attacks work on large language models?},
  author={Duan, Michael and Suri, Anshuman and Mireshghallah, Niloofar and Min, Sewon and Shi, Weijia and Zettlemoyer, Luke and Tsvetkov, Yulia and Choi, Yejin and Evans, David and Hajishirzi, Hannaneh},
  journal={arXiv preprint arXiv:2402.07841},
  year={2024}
}

@inproceedings{galli2024noisy,
  title={Noisy Neighbors: Efficient membership inference attacks against LLMs},
  author={Galli, Filippo and Melis, Luca and Cucinotta, Tommaso},
  booktitle={Proceedings of the Fifth Workshop on Privacy in Natural Language Processing},
  pages={1--6},
  year={2024}
}

@article{wu2025membership,
  title={Membership inference attacks on large-scale models: A survey},
  author={Wu, Hengyu and Cao, Yang},
  journal={arXiv preprint arXiv:2503.19338},
  year={2025}
}

@inproceedings{zhou2025don,
  title={Don’t say no: Jailbreaking llm by suppressing refusal},
  author={Zhou, Yukai and Lou, Jian and Huang, Zhijie and Qin, Zhan and Yang, Sibei and Wang, Wenjie},
  booktitle={Findings ACL'25},
  pages={25224--25249},
  year={2025}
}

@inproceedings{russinovich2025great,
  title={Great, now write an article about that: The crescendo $\{$Multi-Turn$\}$$\{$LLM$\}$ jailbreak attack},
  author={Russinovich, Mark and Salem, Ahmed and Eldan, Ronen},
  booktitle={34th USENIX Security Symposium (USENIX Security 25)},
  pages={2421--2440},
  year={2025}
}

@article{wei2023jailbroken,
  title={Jailbroken: How does llm safety training fail?},
  author={Wei, Alexander and Haghtalab, Nika and Steinhardt, Jacob},
  journal={Advances in Neural Information Processing Systems},
  volume={36},
  pages={80079--80110},
  year={2023}
}

@inproceedings{naseh2025riddle,
  title={Riddle me this! stealthy membership inference for retrieval-augmented generation},
  author={Naseh, Ali and Peng, Yuefeng and Suri, Anshuman and Chaudhari, Harsh and Oprea, Alina and Houmansadr, Amir},
  booktitle={Proc SIGSAC'25},
  pages={1245--1259},
  year={2025}
}

@inproceedings{liu2025mask,
  title={Mask-based membership inference attacks for retrieval-augmented generation},
  author={Liu, Mingrui and Zhang, Sixiao and Long, Cheng},
  booktitle={Proc ACM WebConf'25},
  pages={2894--2907},
  year={2025}
}

@inproceedings{li2025generating,
  title={Generating is believing: Membership inference attacks against retrieval-augmented generation},
  author={Li, Yuying and Liu, Gaoyang and Wang, Chen and Yang, Yang},
  booktitle={ICASSP 2025-2025 IEEE International Conference on Acoustics, Speech and Signal Processing (ICASSP)},
  pages={1--5},
  year={2025},
  organization={IEEE}
}

@inproceedings{anderson2025my,
  title={Is My Data in Your Retrieval Database? Membership Inference Attacks Against Retrieval Augmented Generation},
  author={Anderson, Maya and Amit, Guy and Goldsteen, Abigail},
  booktitle={International Conference on Information Systems Security and Privacy},
  volume={2},
  pages={474--485},
  year={2025},
  organization={Science and Technology Publications, Lda}
}

@inproceedings{mass2022conversational,
  title={Conversational search with mixed-initiative-asking good clarification questions backed-up by passage retrieval},
  author={Mass, Yosi and Cohen, Doron and Yehudai, Asaf and Konopnicki, David},
  booktitle={Proceedings of the Second DialDoc Workshop on Document-grounded Dialogue and Conversational Question Answering},
  pages={65--71},
  year={2022}
}

@inproceedings{sekulic2021towards,
  title={Towards facet-driven generation of clarifying questions for conversational search},
  author={Sekuli{\'c}, Ivan and Aliannejadi, Mohammad and Crestani, Fabio},
  booktitle={Proc. SIGIR'21},
  pages={167--175},
  year={2021}
}

@article{buckley1995automatic,
  title={Automatic query expansion using SMART: TREC 3},
  author={Buckley, Chris and Salton, Gerard and Allan, James and Singhal, Amit},
  journal={NIST special publication sp},
  pages={69--69},
  year={1995},
  publisher={NATIONAL INSTIUTE OF STANDARDS \& TECHNOLOGY}
}

@inproceedings{cao2008selecting,
  title={Selecting good expansion terms for pseudo-relevance feedback},
  author={Cao, Guihong and Nie, Jian-Yun and Gao, Jianfeng and Robertson, Stephen},
  booktitle={Proc SIGIR'08},
  pages={243--250},
  year={2008}
}

@inproceedings{srinivasan2022quill,
  title={QUILL: Query intent with large language models using retrieval augmentation and multi-stage distillation},
  author={Srinivasan, Krishna and Raman, Karthik and Samanta, Anupam and Liao, Lingrui and Bertelli, Luca and Bendersky, Michael},
  booktitle={Proc EMNLP'22 Industry Track},
  pages={492--501},
  year={2022}
}

@inproceedings{zhang2020query,
  title={Query understanding via intent description generation},
  author={Zhang, Ruqing and Guo, Jiafeng and Fan, Yixing and Lan, Yanyan and Cheng, Xueqi},
  booktitle={Proc CIKM'20},
  pages={1823--1832},
  year={2020}
}

@article{marchionini2006exploratory,
  title={Exploratory search: from finding to understanding},
  author={Marchionini, Gary},
  journal={Communications of the ACM},
  volume={49},
  number={4},
  pages={41--46},
  year={2006},
  publisher={ACM New York, NY, USA}
}

@article{belkin1980anomalous,
  title={Anomalous states of knowledge as a basis for information retrieval},
  author={Belkin, Nicholas J},
  journal={Canadian journal of information science},
  volume={5},
  number={1},
  pages={133--143},
  year={1980}
}

@techreport{taylor1967question,
  title={QUESTION-NEGOTIATION AN INFORMATION-SEEKING IN LIBRARIES.},
  author={Taylor, Robert S},
  year={1967}
}

@article{kuhlthau1991inside,
  title={Inside the search process: Information seeking from the user's perspective},
  author={Kuhlthau, Carol C},
  journal={Journal of the American society for information science},
  volume={42},
  number={5},
  pages={361--371},
  year={1991},
  publisher={Wiley Online Library}
}

@article{bates1989design,
  title={The design of browsing and berrypicking techniques for the online search interface},
  author={Bates, Marcia J},
  journal={Online review},
  volume={13},
  number={5},
  pages={407--424},
  year={1989},
  publisher={MCB UP Ltd}
}

@article{sweeney2002k,
  title={k-anonymity: A model for protecting privacy},
  author={Sweeney, Latanya},
  journal={International journal of uncertainty, fuzziness and knowledge-based systems},
  volume={10},
  number={05},
  pages={557--570},
  year={2002},
  publisher={World Scientific}
}

@inproceedings{dwork2006differential,
  title={Differential privacy},
  author={Dwork, Cynthia},
  booktitle={International colloquium on automata, languages, and programming},
  pages={1--12},
  year={2006},
  organization={Springer}
}

@inproceedings{sayed2020test,
  title={A test collection for relevance and sensitivity},
  author={Sayed, Mahmoud F and Cox, William and Rivera, Jonah Lynn and Christian-Lamb, Caitlin and Iqbal, Modassir and Oard, Douglas W and Shilton, Katie},
  booktitle={Proc SIGIR'20},
  pages={1605--1608},
  year={2020}
}

@article{mckechnie2024sara,
  title={SARA: A Collection of Sensitivity-Aware Relevance Assessments},
  author={McKechnie, Jack and McDonald, Graham},
  journal={arXiv preprint arXiv:2401.05144},
  year={2024}
}

@inproceedings{mcdonald2021framework,
  title={A framework for technology-assisted sensitivity review: using sensitivity classification to prioritise documents for review},
  author={McDonald, Graham},
  booktitle={ACM SIGIR Forum},
  volume={53},
  number={1},
  pages={42--43},
  year={2021},
  organization={ACM New York, NY, USA}
}

@inproceedings{mcdonald2018active,
  title={Active learning strategies for technology assisted sensitivity review},
  author={McDonald, Graham and Macdonald, Craig and Ounis, Iadh},
  booktitle={Advances ECIR'18},
  pages={439--453},
  year={2018},
  organization={Springer}
}

@article{vinjumur2015finding,
  title={Finding the privileged few: Supporting privilege review for e-discovery},
  author={Vinjumur, Jyothi K and Oard, Douglas W},
  journal={Proceedings of the Association for Information Science and Technology},
  volume={52},
  number={1},
  pages={1--4},
  year={2015},
  publisher={Wiley Online Library}
}

@article{oard2018jointly,
  title={Jointly minimizing the expected costs of review for responsiveness and privilege in e-discovery},
  author={Oard, Douglas W and Sebastiani, Fabrizio and Vinjumur, Jyothi K},
  journal={ACM Transactions on Information Systems (TOIS)},
  volume={37},
  number={1},
  pages={1--35},
  year={2018},
  publisher={ACM New York, NY, USA}
}

@inproceedings{mckechnie2024bi,
  title={Bi-Objective Negative Sampling for Sensitivity-Aware Search},
  author={McKechnie, Jack and McDonald, Graham and Macdonald, Craig},
  booktitle={Proc SIGIR'24},
  pages={2296--2300},
  year={2024}
}

@inproceedings{mckechnie2024cascading,
  title={Cascading Ranking Pipelines for Sensitivity-Aware Search},
  author={McKechnie, Jack},
  booktitle={Advances ECIR'18},
  pages={331--333},
  year={2024},
  organization={Springer}
}

@inproceedings{olteanu2021facts,
  title={FACTS-IR: fairness, accountability, confidentiality, transparency, and safety in information retrieval},
  author={Olteanu, Alexandra and Garcia-Gathright, Jean and de Rijke, Maarten and Ekstrand, Michael D and Roegiest, Adam and Lipani, Aldo and Beutel, Alex and Olteanu, Alexandra and Lucic, Ana and Stoica, Ana-Andreea and others},
  booktitle={ACM SIGIR Forum},
  volume={53},
  number={2},
  pages={20--43},
  year={2021},
  organization={ACM New York, NY, USA}
}

@article{sanchez2016c,
  title={C-sanitized: A privacy model for document redaction and sanitization},
  author={S{\'a}nchez, David and Batet, Montserrat},
  journal={Journal of the Association for Information Science and Technology},
  volume={67},
  number={1},
  pages={148--163},
  year={2016},
  publisher={Wiley Online Library}
}

@article{anandan2012t,
  title={t-Plausibility: Generalizing words to desensitize text.},
  author={Anandan, Balamurugan and Clifton, Chris and Jiang, Wei and Murugesan, Mummoorthy and Pastrana-Camacho, Pedro and Si, Luo},
  journal={Trans. Data Priv.},
  volume={5},
  number={3},
  pages={505--534},
  year={2012}
}

@inproceedings{baron2023using,
  title={Using ChatGPT for the FOIA Exemption 5 Deliberative Process Privilege.},
  author={Baron, Jason R and Rollings, Nathaniel W and Oard, Douglas W},
  booktitle={LegalAIIA@ ICAIL},
  pages={32--48},
  year={2023}
}

@article{krasakis2024corpus,
  title={Corpus-informed Retrieval Augmented Generation of Clarifying Questions},
  author={Krasakis, Antonios Minas and Yates, Andrew and Kanoulas, Evangelos},
  journal={arXiv preprint arXiv:2409.18575},
  year={2024}
}
\end{document}